\documentstyle[eqsecnum,pre,preprint,aps,epsfig]{revtex}

\tightenlines

\newcommand{\bq}{\begin{equation}}
\newcommand{\eq}{\end{equation}}
\newcommand{\bqa}{\begin{eqnarray}}
\newcommand{\eqa}{\end{eqnarray}}
\newcommand{\ben}{\begin{enumerate}}
\newcommand{\een}{\end{enumerate}}
\newcommand{\bc}{\begin{center}}
\newcommand{\ec}{\end{center}}
\newcommand{\bqb}{\begin{eqnarray*}}
\newcommand{\eqb}{\end{eqnarray*}}

\begin{document}

\draft
\preprint{PM/04-53,~~December 2004}

\title{  Special Supersymmetric features of large invariant mass unpolarized and
polarized top-antitop production at LHC.
\footnote{Partially supported by EU contract HPRN-CT-2000-00149}}
\author{M. Beccaria$^{a,b}$,
S.Bentvelsen$^c$, M.Cobal$^d$,
F.M. Renard$^e$ and C. Verzegnassi$^{f, g}$ \\
\vspace{0.4cm}
}

\address{
$^a$Dipartimento di Fisica, Universit\`a di
Lecce \\
Via Arnesano, 73100 Lecce, Italy.\\
\vspace{0.2cm}
$^b$INFN, Sezione di Lecce\\
\vspace{0.2cm}
$^c$ Nikhef, Amsterdam, The Netherlands\\
\vspace{0.2cm}
$^d$ Dipartimento di Fisica, Universita' di Udine, V. delle 
Scienze 208,\\ Udine and INFN Sezione di Trieste \\
\vspace{0.2cm}
$^e$ Physique
Math\'{e}matique et Th\'{e}orique, UMR 5825\\
Universit\'{e} Montpellier
II,  F-34095 Montpellier Cedex 5.\hspace{2.2cm}\\
\vspace{0.2cm}
$^f$
Dipartimento di Fisica Teorica, Universit\`a di Trieste, \\
Strada Costiera
 14, Miramare (Trieste) \\
\vspace{0.2cm}
$^g$ INFN, Sezione di Trieste\\
}

\maketitle

\begin{abstract}

\vspace{-1cm}

We consider the top-antitop invariant mass distributions for 
production of
unpolarized and polarized top quark pairs at LHC, in the theoretical
framework of the MSSM. Assuming a "moderately" light SUSY scenario, we
derive the leading logarithmic electroweak contributions at one 
loop in a
region of large invariant mass, $M_{t\bar t}\simeq1$ TeV, for the unpolarized
differential cross section $d\sigma/dM_{t\bar t}$ and for the differential
longitudinal top polarization asymmetry $A_t(M_{t\bar t})$. 
We perform a realistic evaluation
of the  expected uncertainties of the two
quantities, both from a theoretical and from an experimental point
of view, and discuss the possibility of obtaining, from accurate
measurements  of the two mass distributions, stringent 
consistency tests of the  model, in particular identifications of large $\tan\beta$ effects.

\end{abstract}

\section{Introduction}

The relevance of top quark physics at LHC is nowadays firmly
established. Exhaustive detailed descriptions of the experimental
strategies  that will be adopted to determine with the greatest obtainable
precision the observable properties of the measurable processes are
available in the literature \cite{CERNYB}. 
Within the Standard
Model (SM), one sees from a careful reading of Ref.\cite{CERNYB} that top 
production will
improve our knowledge of fundamental features of the model 
that are still
relatively poorly determined, like the precise value of the 
top mass (aimed
LHC accuracy of 1 GeV) or that of the CKM $V_{tb}$ coupling 
(aimed 5 \% precision). For physics beyond the SM, top production offers 
a variety of
possible tests that are also fully listed in Ref.\cite{CERNYB}.
Generally speaking,
these assume the possibility of detecting small effects, typically due to
virtual exchanges of particles not existing in the SM. In order to be able
to identify unambiguously such terms, a suitable control of all possible
(theoretical and experimental) uncertainties is therefore essential. 
This
requires dedicated analyses and accurate calculations from all the
involved research groups. \par
On the theoretical side, a widespread hope exists that LHC will produce a
certain amount of supersymmetric particles. Assuming that this will be the
case, top quark physics might then be able to provide consistency tests of
the candidate supersymmetric model via identification of small
supersymmetric virtual effects, provided that the latter were sufficiently
larger than the intrinsic overall uncertainty that would affect the
measured process.
In the specific framework of the Minimal Supersymmetric
Standard Model, the calculation of the virtual electroweak effects has
been actually performed at the perturbative one-loop level for arbitrary
initial partons c.m.energy values. In a first paper \cite{Hollik} the
production of unpolarized top quarks has been considered. 
In general, the
virtual electroweak SUSY effects can become appreciably large, 
reaching,
roughly, the ten percent size,
depending on the values of the several (six,
in the approach of Ref.\cite{Hollik}) relevant theoretical parameters of the Model. To
appreciate and identify these supersymmetric contributions, the
theoretical uncertainty of the production process, essentially of QCD
origin, should be therefore controlled and maintained below the 10 \%
level. This drastic operation might be completely avoided if the measured
quantity were the QCD-free final top longitudinal polarization asymmetry,
already considered for LHC measurements in a previous paper\cite{Kao}. 
In particular, from Ref.\cite{Kao} one
realizes the existence of a strong dependence of the differential
asymmetry at LHC on $\tan\beta$, for given values of other supersymmetric
parameters. Typically, the magnitude of the asymmetry lies in the few
percent range. To appreciate it, experimental measurements at  this
precision level would be therefore requested.\par
The theoretical analyses of Refs.\cite{Hollik}, \cite{Kao}
were performed for arbitrary
initial parton c.m. energy values. Under these conditions, all the
parameters of the model enter, in principle, in the theoretical
expressions. To reduce the number of parameters that effectively determine
a prediction would be a welcome feature for any accurate test. In the
specific case of the MSSM, it has been recently shown 
\cite{HQ}, that a great simplification can be achieved when the
final top-antitop pair is produced at a c.m. energy $\sqrt{s}$ 
sufficiently
larger than the heaviest mass $M$ of the SUSY particles that contribute via
virtual exchanges. Assuming a moderately light SUSY scenario, e.g.
$M\simeq350$ GeV, imposes a qualitative range $\sqrt{s}\simeq 1$ TeV 
(i.e.
$s/M^2 \simeq10$). Under these conditions, an asymptotic logarithmic
expansion at one loop of so called Sudakov type can be used to provide a
satisfactory description of the differential cross 
sections \cite{Sud}. To the next-to leading linear logarithmic order
accuracy, the only supersymmetric parameter that enters the expansion is
$\tan\beta=v_2/v_1$, via supersymmetric effects of Yukawa kind that only affect
the linear logarithmic term \cite{Sud}.
All the remaining SUSY parameters are
shifted in the next terms of the expansion, 
and in this preliminary analysis we shall neglect them on the basis of previous
studies performed for LC physics ~\cite{LC}, devoting a more rigorous discussion
of their possible effects at LHC in the final conclusions.

The aim of this paper is that of investigating which useful information on
the MSSM would be derivable, in the moderately light SUSY scenario that we
described (or in an essentially similar one), by accurate measurements of
the large (about 1 TeV) invariant mass distributions of the unpolarized
top-antitop production cross section and of the aforementioned final top
longitudinal polarization asymmetry. With this aim, we shall organize the
paper  in the following way. In Section 2, we shall quickly exhibit the
relevant asymptotic expansions for polarized and unpolarized top
production, working in the c.m. frame of the elementary initial partons
(essentially, gluon-gluon frame), and show the special and simple
dependence on $\tan\beta$ that appears, discussing qualitatively its possible
virtues, particularly in the large ($\tan\beta>10$) values
region. In the following Section 3 the expected theoretical and experimental uncertainties
will be discussed, starting from the very initial request of transforming
the initial partons c.m. energy $\sqrt{s}$-dependent predictions into measurable
({\em i.e.} top-antitop invariant mass-dependent) expressions
and making a conservative estimate of the reasonable experimental accuracies
obtainable in the measurements. Section 4 will show which possible
information on the MSSM would be derivable under the previous assumptions,
providing a final discussion with a number of possibly general
conclusions.

\section{DIFFERENTIAL CROSS SECTIONS AT PARTONIC LEVEL}

In this Section, we shall perform the preliminary logarithmic expansion of
the relevant differential
cross sections at  the partonic level, for production of a polarized
top-antitop pair, in the c.m. system
of the initial parton pair. As we anticipated, we shall assume a
preliminary production of supersymmetric
particles, and a moderately light SUSY scenario where all supersymmetric
masses lie below, say,
350 GeV. Under this assumption, a reasonable choice for the  validity of
an asymptotic expansion in
the c.m. energy $\sqrt{s}$ seems to us to be $\sqrt{s}\simeq 1$ TeV, 
so that, denoting
with $M$ the
mass of the heaviest supersymmetric  particle that provides virtual
contributions to the process,
$s/M^2 \simeq 10$ (for reasonably different values of $M$, this qualitative
request fixes the corresponding value
of $\sqrt{s}$).\par
The choice of $\sqrt{s}$  generates useful simplifications in the theoretical
description of the process, that have
 already been stressed in a previous paper \cite{HQ}. 
In fact, for production at
 LHC, one already knows that the largely dominant contributions come from
the initial gluon-gluon
state. At the Born level, the corresponding scattering amplitude is given
by the sum of three
 $s$, $t$, $u$ channel diagrams. When $\sqrt{s}$ is much larger than the top mass, only
the two $t$, $u$ channels contribution
for opposite helicity gluons survives, while the sum of the $s$,$t$,$u$ channel
diagrams with equal helicity
gluons vanishes as $m^2_t/s$ . An analogous simplification appears
for the final top-antitop pair
since the final top helicity $\lambda\equiv\lambda_t$ is opposite 
to the antitop one $\lambda_{\bar t}$, neglecting $m^2_t/s$ terms;
note that this corresponds to chirality conservation along the
fermionic line since, for the antitop, chirality is opposite
to helicity. As a consequence of these
facts, we shall concentrate our attention, at the electroweak one-loop
order  that we shall consider, on the MSSM corrections to the  gluon-gluon
$t$,$u$ channel Born diagrams for opposite top-antitop helicities (actually,
in Ref.\cite{HQ} a detailed calculation of the one-loop corrections
to the initial $q\bar q$ parton state  was also performed and shown, and the
numerical overall calculation confirmed the large dominance of the two
gluon channel).\par
In the chosen picture, the scattering amplitude for the process is, as we
said, only given by the sum of the $t$,$u$
channel terms

\bq
A^{Born, t}= -~{g^2_s\over t}
[\bar
u(t){\lambda^i\lambda^j\over4}(\gamma^{\mu}\epsilon^i_{\mu})
(\gamma^{\rho}(k^i-p^{q'})_{\rho})(\gamma^{\nu}\epsilon^j_{\nu})
 v(\bar t)]
\label{bornt}\eq

\bq
A^{Born, u}= -~{g^2_s\over u}
[\bar
u(t){\lambda^j\lambda^i\over4}(\gamma^{\mu}\epsilon^j_{\mu})
(\gamma^{\rho}(k^i-p^{\bar q'})_{\rho})(\gamma^{\nu}\epsilon^i_{\nu})
 v(\bar t)]
\label{bornu}
\eq
\noindent
whose contribution to the helicity amplitude
$F^{Born}(\mu,-\mu,\lambda,-\lambda)$, where $\mu$ and $\lambda$
denote the gluon and top helicities, is :

\bq
F^{Born}(\mu,-\mu,\lambda,-\lambda)
=g^2_s({\lambda^i\lambda^j\over4})
{2\lambda\cos\theta+\mu\over1-\cos\theta}\sin\theta
+g^2_s({\lambda^j\lambda^i\over4})
{2\lambda\cos\theta+\mu\over1+\cos\theta}\sin\theta
\label{fborn}\eq
\noindent
where $g_s$ is the QCD coupling constant, $i$ and $j$ refer to the two gluon color
states, $\lambda^i$ and $\lambda^j$ are color matrices,  $\epsilon^i_\nu$ 
and  $\epsilon^j_\nu$ the polarization vectors, $\theta$ is the angle between the top quark
and the gluon.

The corresponding expression of the differential cross sections for fixed
top helicity ($L$, $R$ denote the chirality of both top and antitop) 
are then :

\bq
{d\sigma^{Born}(gg\to t_L\bar t_L)\over
dcos\theta}=
{d\sigma^{Born}(gg\to t_R\bar t_R)\over
dcos\theta}={\pi\alpha^2_s\over 8s}[ {u^2+t^2\over 3ut}-
{3(u^2+t^2)\over 4s^2}]
\label{sigborn}\eq

The electroweak corrections turn out to be extremely simple at the
one-loop level. In the adopted
logarithmic expansion of Sudakov type, and using the conventional
definitions of the various terms
that one can find {\em e.g.} in Ref.\cite{HQ}, only universal gauge and
Yukawa terms appear. Graphically, they are related to the diagrams shown
in Fig.1. Briefly, their contribution to the differential cross sections
can be
expressed by the following simple equations :

\bqa
&&{d\sigma^{1~loop}(gg\to t_L\bar t_L)\over
d\cos\theta}={d\sigma^{Born}(gg\to t_L\bar t_L)\over
d\cos\theta}(1+2c^{t\bar t}_{L})\nonumber\\
&&={\pi\alpha^2_s\over 8s}[ {u^2+t^2\over 3ut}-
{3(u^2+t^2)\over 4s^2}]~[1+ \nonumber \\
&& \frac{\alpha}{72\pi s_{W}^2 c_{W}^2}(27-26s_{W}^2)(2\log\frac{s}{M^2_W}-\log^2\frac{s}{M^2_W}) \nonumber\\
&& -\frac{\alpha}{4M_W^2\pi s_{W}^2}(m_t^2(1+\cot^2\beta)+m_b^2(1+\tan^2\beta))
\log\frac{s}{M^2_W}] 
\label{sigbornL}\eqa

\bqa
&&{d\sigma^{1~loop}(gg\to t_R\bar t_R)\over
d\cos\theta}={d\sigma^{Born}(gg\to t_R\bar t_R)\over
d\cos\theta}(1+2c^{t\bar t}_{R})\nonumber\\
&&={\pi\alpha^2_s\over 8s}[ {u^2+t^2\over 3ut}-
{3(u^2+t^2)\over 4s^2}]~[1+\nonumber\\
&& \frac{2\alpha}{9\pi c_{W}^2}(2\log\frac{s}{M^2_W}-\log^2\frac{s}{M^2_W}) 
-\frac{\alpha}{2M_W^2\pi s_{W}^2} m_t^2(1+\cot^2\beta)\ \log\frac{s}{M^2_W}]
 \label{sigbornR}\eqa
using the universal coefficients $c^{t\bar t}_{L,R}$
defined in \cite{HQ,Sud}.

Eqs.\ref{sigbornL},\ref{sigbornR} are only intermediate expressions that allow to derive the two
quantities that we shall consider in what follows. The first one is the
(usual) unpolarized top-antitop cross section, that in the chosen scenario
reads :

\bqa
&&{\frac{d\sigma^{1~loop}_U}{
d\cos\theta} \stackrel{def}{=} \frac{d\sigma^{1~loop}(gg\to t_L\bar t_L +t_R\bar t_R)}
{dcos\theta}}={d\sigma^{Born}(gg\to t_L\bar t_L +t_R\bar t_R)\over
dcos\theta}(1+c^{t\bar t}_{L}+c^{t\bar t}_{R}) = \nonumber\\
&&={\pi\alpha^2_s\over4s}[ {u^2+t^2\over 3ut}-
{3(u^2+t^2)\over 4s^2}]~[1+{\alpha\over144\pi s^2_Wc^2_W}
(27-10s^2_W)(2\log\frac{s}{M^2_W}-\log^2\frac{s}{M^2_W})\nonumber\\
&&-~{\alpha \over8\pi s^2_W}\log\frac{s}{M^2_W}({3m^2_t\over M^2_W}
(1+\cot^2\beta)+{m^2_b\over M^2_W}(1+\tan^2\beta))]
\label{sig1}\eqa

 The second quantity that we shall consider in this paper will be called
the final top longitudinal polarization asymmetry and denoted by us with
the symbol :

\bqa
a_t(\theta)&=&[{d\sigma^{1~loop}(gg\to t_L\bar t_L)\over d\cos\theta}
-{d\sigma^{1~loop}(gg\to t_R\bar t_R)\over d\cos\theta}]
/ \nonumber \\
&& [{d\sigma^{1~loop}(gg\to t_L\bar t_L)\over dcos\theta}
+{d\sigma^{1~loop}(gg\to t_R\bar t_R)\over dcos\theta}]
\label{at}
\eqa

At Born level, it is vanishing. Its theoretical expression 
at first order in the chosen configuration is :

\bqa
a_t &\simeq&c^{t\bar t}_{L}-c^{t\bar
t}_{R}=
{\alpha(9-14s^2_W)\over48\pi s^2_Wc^2_W}
[2\log\frac{s}{M^2_W}-\log^2\frac{s}{M^2_W}]
\nonumber\\
&&-{\alpha\over8\pi s^2_WM^2_W}[\log\frac{s}{M^2_W}][m^2_b(1+\tan^2\beta)
-m^2_t((1+\cot^2\beta)]
\label{at1}\eqa

It should be mentioned that we have considered a quantity already defined
in  previous papers, in particular for LHC by Kao and Wackeroth
\cite{Kao}. Our
definition differs by their definition only by a sign.
The reason why we consider this asymmetry a potentially interesting
observable has been already indicated in Ref.\cite{Kao}, since at this
perturbative order $a_t$ is completely free of QCD effects (and related
theoretical uncertainties), and we shall come back to this point in the
next Section.\par

In Eqs. 2.5-2.7,2.9 we have given Sudakov expressions for the one-loop corrections
in the MSSM case. The corresponding expressions in the SM case can be obtained
in a straightforward way by the replacement~\cite{Sud}: $2\log-\log^2$
$\to$ $3\log-\log^2$, $2m_t^2(1+\cot^2\beta)\to m_t^2$ and $2m_b^2(1+\tan^2\beta)\to m_b^2$.

An important point that we should now clarify is in fact our
treatment of the QCD corrections to the considered process. Generally
speaking they are of two natures, those of SM origin and the extra
ones of SUSY nature. Since the aim of this paper is the identification
of genuine supersymmetric effects, the SM QCD effects will be
considered by us as "partially" known terms, whose theoretical
uncertainty will be discussed in Section 3. The treatment of the SUSY
QCD corrections will be different. For these terms, another benefit of
the large $\sqrt{s}$ choice is the disappearance (at the one loop
order) of the correction to the $s$-channel diagram, containing the SUSY
contributions to the gluon self-energy. The remaining term arises from
the $t$, $u$ channel vertex effects with gluino exchanges. In our scenario
which assumes a gluino mass smaller than about 350 GeV, its Sudakov
logarithmic (linear) contribution to the scattering amplitude is given
by the following expression:

\bq
-F^{Born}(\mu,-\mu,\lambda,-\lambda) \ {\alpha_s\over3\pi}\log{s\over M^2}
\label{coefewqcd}\eq

Its effect in the unpolarized differential cross section can be easily
computed, and adds a contribution that turns out to be, numerically,
smaller than that of electroweak origin in our scenario. As stressed
in Ref.\cite{HQ} this surviving contribution has the same negative
sign as that of the electroweak one, which enhances the SUSY effect,
and a numerical size of a few percent in the considered $\sqrt{s}$
region, which makes its one loop expression reasonably safe. On the
contrary, the suppressed $s$-channel contribution would have been
positive, thus decreasing the overall SUSY effect.\par

Until now, our analysis has been limited to the description of the
elementary partonic process, in particular of
its electroweak properties. Clearly, this is only a first step towards a
realistic prediction for observable   quantities, that fully takes into
account the effective nature of the initial and of the final states.
This
will be tentatively done in the next two Sections. But before entering
this concrete stage, we will illustrate with
a few extra analyses at the partonic level the reasons why we believe that
suitable measurements of top-antitop
production in our proposed scenario might be interesting. With this aim,
we shall retain for the moment as variable quantity of the process the
initial gluon pair c.m. energy $\sqrt{s}$ (whose relationship with the
observable invariant mass $M_{tt}$ will be examined in Section 3) 
and assume
that an experimental measurement can be performed for 
two proposed $\sqrt{s}$
distributions. The first one is :

\bqa
{d\sigma(PP\to t\bar t+...)\over ds}&=&
{1\over S}~\int^{\cos\theta_{max}}_{\cos\theta_{min}}
d\cos\theta~[~\sum_{ij}~L_{ij}(\tau, \cos\theta)
{d\sigma_{ij\to  t\bar t}\over d\cos\theta}(s)~]
\eqa
\noindent
where $\tau={s\over S}$, and $(ij)$ represent 
all initial $q\bar q$ pairs with 
$q=u,d,s,c,b$ and the initial $gg$ pairs, with the corresponding
luminosities

\bq
L_{ij}(\tau, \cos\theta)={1\over1+\delta_{ij}}
\int^{\bar y_{max}}_{\bar y_{min}}d\bar y~ 
~[~ i(x) j({\tau\over x})+j(x)i({\tau\over x})~]
\eq
\noindent
where S is the total pp c.m. energy, and 
$i(x)$ the distributions of the parton $i$ inside the proton
with a momentum fraction,
$x={\sqrt{s\over S}}~e^{\bar y}$, related to the rapidity
$\bar y$ of the $t\bar t$ system~\cite{QCDcoll}.
The parton distribution functions are the 2003 NLO MRST 
set available on~\cite{wlumi}.
The limits of integrations for $\bar y$ can be written

\bqa
&&\bar y_{max}=\max\{0, \min\{Y-{1\over2}\log\chi,~Y+{1\over2}\log\chi,
~-\log(\sqrt{\tau})\}\}\nonumber\\
&&
\bar y_{min}= - \bar y_{max}
\eqa
\noindent
where the maximal rapidity is $Y=2$, the  
quantity $\chi$ is related to the scattering angle
in the $t\bar t$ c.m.
\bq
\chi={1+\cos\theta\over1-\cos\theta} 
\eq
and 
\bq
\cos\theta_{min,max}=\mp\sqrt{1-{4p^2_{T,min}\over s}}
\eq
expressed in terms of
the chosen value for $p_{T,min} = 10$ GeV.\\

In an analogous way, we define :

\bq
\label{def:at}
A_t(s)=[{d\sigma_{L}\over ds}
-{d\sigma_{R}\over ds}]/[{d\sigma_{L}\over ds}
+{d\sigma_{R}\over ds}]
\eq
with

\bqa
{d\sigma_{L,R}\over ds}&=&
{1\over S}~\int^{\cos\theta_{max}}_{\cos\theta_{min}}
d\cos\theta~[~\sum_{ij}~L_{ij}(\tau, \cos\theta)
{d\sigma_{ij\to  t_{L,R}\bar t_{L,R}}\over d\cos\theta}(s)~]
\eqa

We have drawn in Figs.\ref{fig2} and \ref{fig3} the  values of the two distributions versus
the c.m energy $\sqrt{s}$ in the energy
range that would be relevant in our scenario, from roughly 700 GeV to
roughly 1 TeV, for three significant values of  $\tan\beta$ i.e. 
$\tan\beta=1,10,50$. 
These have been obtained integrating our expressions of the two
gluon process  written in this Section and adding to them the extra
depressed initial quark-squark terms, given in Ref.\cite{HQ} and not
rewritten here for simplicity purposes. 
As one sees, $A_t(s)$ changes sign in the range. 
It should be stressed that our values of this
asymmetry are in excellent agreement with the previous results of
Ref.\cite{Kao} (Fig.3 of that paper) if the comparison 
is limited to the set of
parameters of that Reference that would correspond to our scenario
(i.e. those corresponding to the smallest SUSY masses). 
One sees also that the electroweak MSSM effect
could be relatively large. In $d\sigma/ds$ 
it would reach a relative 20-22 \%
for $\tan\beta\simeq 50$ and $\sqrt{s}=1$ TeV ; in $A_t(s)$, the effect is smaller, of the six
percent size for the same values of the parameters.  
One should keep in mind, though, that it represents in this
case, not a small correction to a main SM value, but essentially the
overall value of the asymmetry (which would be at the one percent level in the SM).\par
An additional feature of the two considered distributions is that the dependence on $\tan\beta$
is rather strong. This can be understood by looking at Eqs.(2.7-2.9) 
and is graphically shown in Figs.\ref{fig4}-\ref{fig5}. One notices in particular an enhancement
of the effect for large $\tan\beta$ values in both observables. In the 
asymmetry the effect changes sign when $\tan\beta$ moves from the low ($\simeq 1$) values
to the large ($\simeq 50$) ones, a fact that can also be easily understood looking at 
Eq.2.9.

Encouraged by these promising  features, we considered at this point at a qualitative level
the possibility 
of deriving confidence limits on $\tan\beta$ from a $\chi^2$ analysis of the two observables.
With this purpose, we have used the expected event distribution shown in Fig.\ref{fig6} 
and taken from~\cite{Cogneras}.
It  has been obtained from the semileptonic decay channel into muons only. The integrated luminosity,
taking into account the branching ratio, is of about 5 fb$^{-1}$, corresponding to half a  year of data taking 
at low luminosity ($10^{33} {\rm cm}^2 {\rm s}^{-1}$).
This has provided us with a realistic statistical experimental error and for this analysis we have not
considered the other sources of uncertainties, that will be discussed in the next Section 3. Note
that the event distribution that we have used is given as a function of the top-antitop invariant
mass $M_{t\bar t}$. In this qualitative analysis, we have assumed the equality $M_{t\bar t} = \sqrt{s}$,
whose validity will also be discussed in the next Section 3.

In agreement with the spirit of this preliminary paper, we shall retain, we repeat, 
the two (leading and next to leading) terms of the logarithmic expansion in the analysis.
A discussion of the possible effects of next to next to leading terms will be given in the final 
conclusions.

The results of our $\chi^2$ fits to $\tan\beta$ are shown in Fig.\ref{fig7}, in the two cases of 
(a) measurement of only $d\sigma_U/ds$ and (b) measurement of both $d\sigma_U/ds$ and $A_t$.
As one sees from these figures, it would be possible to obtain from our analysis the following main 
information: 
\begin{enumerate}
\item A measurement of only  $d\sigma_U/ds$ would provide information in two separate
ranges, $1 \lesssim \tan\beta \lesssim 3$ and $20 \lesssim \tan\beta$. This can be understood as follows.
In our approximation, the unpolarized cross section depends on $\tan\beta$ through the precise Yukawa
correction which is a definite combination of $\tan^2\beta$ and $\cot^2\beta$. Each value of the correction 
is obtained from a couple of {\em mirror} values of $\tan\beta$. A $\chi^2$ analysis is thus plagued 
by this irreducible uncertainty and will always show a false minimum at the mirror value of the {\em true}
$\tan\beta$. However, if $1 \lesssim \tan\beta \lesssim 3$ or $20 \lesssim \tan\beta$, there is 
a  $\Delta\chi^2=1$
confidence region centered around the exact value of $\tan\beta$ and we have drawn in the Fig.\ref{fig7}
the corresponding boundaries. If, on the other hand, $\tan\beta$ falls in the {\em dead region} 
$3 \lesssim \tan\beta \lesssim 20$, it is practically impossible to tell between the true and false 
(mirror) minima of $\chi^2$ and we have not drawn any line.

\item The additional information obtained by the measurement of $A_t$ is crucial to
eliminate the above mentioned {\em dead region} $3 \lesssim \tan\beta \lesssim 20$. 

\item The more interesting information is, in our opinion, that relative to the large $\tan\beta$ region,
roughly $\tan\beta \gtrsim 20$. In fact, in the low $\tan\beta$ $\simeq 1$ region there should be 
alternative accurate determinations provided by other measurements~\cite{Abdel}. On the
contrary, for the large $\tan\beta$ values, no accurate alternative measurements at the LHC time 
seem to be, to our knowledge, available.

\item In the large $\tan\beta$ region, while the measurements of $d\sigma_U/ds$ alone seems to 
provide already a satisfactory result (with a relative error below 20 \% for $\tan\beta$ larger than
about 30), the addition of $A_t$ would lead to an extremely accurate determination 
reaching a few percent value for $\tan\beta\simeq 50$.
\end{enumerate}

This qualitative analysis is now completed. We feel that the  features
that have emerged justify the effort of
moving to the detailed consideration of the related physical process, and
of  examining whether the nice
properties that we underlined at qualitative level survive in the realistic
experimental measurements. This will be done starting in the forthcoming
Section 3.

\section{Experimental and theoretical uncertainties}

The discussion of virtual supersymmetric effects has led to corrections
to the lowest level of the partonic cross section for top-anti-top
production as function
of the partonic centre-of-mass $\sqrt{s}$. Notably in the high region
of $\sqrt{s}$, the corrections can be as large as 20 to 30\% and
depend on the exact value of the supersymmetric parameters. In this
section we concentrate on the experimental issues involved in order to
measure these corrections. 

Although it can be inferred from Fig.\ref{fig2} that the shape of the
distribution of $M_{tt}$ as function of $\sqrt{s}$ is distorted by the
supersymmetric effects, most of the sensitivity to these new effects
originate in the difference of the $t\bar{t}$ cross section above
$\sqrt{s}$ of approximately 600 GeV.  The determination of absolute
cross sections at hadron machines is however highly non-trivial. The
estimated overall predicted theoretical uncertainty of $t\bar{t}$
production is of the order of 12\%~\cite{CERNYB}, the error coming
from remaining scale dependence of the NLO QCD calculations.

Experimentally there are two main concerns to determine the $t\bar{t}$
cross section as function of $\sqrt{s}$. First, one can only measure
the final state top pairs, and has therefore experimental access to
the invariant mass $M_{tt}$ only. NLO QCD effects (final state gluon
radiation, virtual effects) spoil the equivalence of $M_{tt}$ with
$\sqrt{s}$.  Second, detector effects like efficiency uncertainties, 
jet resolutions and mis-calibrations migrate the true $M_{tt}$
distribution. We discuss now these two effects.

\subsection{Higher order QCD effects}

Higher order QCD effects perturb the determination of the cross
section as function of $\sqrt{\hat{s}}$ in two ways.  The $t\bar{t}$
cross section increases from 590 pb to $830\pm 100$ pb, calculated in
LO and NLO respectively, at the LHC. Besides the normalisation, also
the shape of the $M_{t\bar t}$ distribution gets distorted by NLO
effects, due to real and virtual effects.

The effects of the NLO QCD calculations have been  investigated using the
MC@NLO~\cite{mcatnlo} program, that incorporates a full NLO treatment in
the Herwig MC generator. We have used this program to generate the
$M_{t\bar t}$ distribution both in LO and in NLO, using identical parton
shower, parameters settings etc.  The effect of the NLO calculations
on the shape of the $M_{t\bar t}$ distribution is shown in Fig.~\ref{fig8}, 
where the ratio NLO/LO is given as a function of $M_{t\bar t}$ at NLO. 
The value of $M_{t\bar t}$ is obtained at the parton level, as
the invariant mass of the top and anti-top quark, after both ISR and
FSR.  The LO and NLO total cross sections were normalised to each
other. Deviations from unity in this figure are entirely due to
differences in shape of $M_{t\bar t}$. As one sees from the figure,
the relative difference between $\sqrt{s}$ and $M_{t\bar t}$
remains bounded (below roughly 5\%) when $\sqrt{s}$ varies between 
$\sim 700 $ GeV and 1 TeV, which is the chosen energy range of this paper.
For larger $\sqrt{s}$ values the difference raises and can reach a 10 \%
limit when $\sqrt{s}$ approaches what we consider a realistic experimental
limit (see Fig.~\ref{fig6}) of the search, i.e. $\sqrt{s} \simeq 1.3$ TeV.

\subsection{Experimental systematic uncertainties}

Top events are triggered and selected best when they decay
semileptonically, i.e. where one top decays to a b-jet plus a lepton
and a neutrino, the other top decays to a b-jet and two light quark
initiated jets. The lepton is ideal to `trigger' the event, and the
two b-jets are clear signatures for top production. The kinematics of
the neutrino can be recovered as the missing transverse energy of the
event. Ambiguities in the longitudinal direction can be dealt with by
consistency requirements on the event topology (for example requiring
two tops to have the same mass).  We generated 1.1$\times$10$^6$ $t
\bar t$ events, using the Pythia Monte Carlo
\cite{Pythia} and processing them through the ATLAS detector fast
simulation \cite{Atlfast}. The number of events correspond to about 5
fb$^{-1}$ of collected data. The selection and reconstruction criteria
applied are described in Ref. \cite{Cogneras}, resulting in an
efficiency of about 1.5\%. This overall efficiency results from the
stringent cuts applied to select the events, including the
b-tagging. These requirements were applied to ensure a high purity of
our $t \bar t $ sample, and the surviving events contain a negligible
amount of background not originated from ttbar production.  A first
estimate of the experimental errors involved is obtained by
considering two of the main sources of systematic uncertainties in the
determination of the invariant mass $M_{t\bar t}$; the jet energy
scale uncertainty and the uncertainties of jet development due to
initial and final state showering.  To evaluate the effect of an
absolute jet energy scale uncertainty, a 5\% miscalibration
coefficient was applied to the jet energies.  This produces a
bin-by-bin distortion of the $M_{t \bar t}$ distribution smaller than
$20\%$.  This is certainly an overestimate of the possible error,
since it has been shown that in ATLAS an accurate absolute energy
calibration of light quarks and b-jets can be extracted from Z+jet
events, with an expected precision of about 1\%~\cite{CERNYB}. An
in-situ calibration of light jets in which both the absolute energy
and direction calibration are extracted from the $W \rightarrow jj$
channel itself is possible as well. For this purpose a cleaner sample
of W candidates can be selected from the $t \bar{t}$ events.

Uncertainties in the initial state radiation from the incoming partons
(ISR) and final state radiation from the top decay products (FSR)
affect the precision of the $M_{t \bar t}$ measurement. To estimate
this effect, as suggested in ~\cite{CERNYB}, the $M_{t \bar{t}}$
distribution obtained with the standard Pythia Monte Carlo has been
compared with the same distribution determined with ISR switched
off. The same approach was employed for FSR.  The level of knowledge
of ISR and FSR is of the order of 10$\%$ so the systematic uncertainty
on each bin of the $t \bar{t}$ mass was taken to be of 20\% of the
corresponding difference in the number of events obtained comparing
the standard mass distribution with the distribution having switched
off ISR (or FSR).  This results as well in an error that is smaller
than 20 \%.

Experimentally, the
luminosity introduces an ultimate uncertainty of the order of 5\%; at
the startup period of LHC this uncertainty will be much larger.
In conclusion, from a rather conservative evaluation of the most relevant,
theoretical and experimental, uncertainties, we are led to the 
conclusion that an overall error of approximately 20 \%- 25 \% in our
energy range appears realistically achieavable from a 
preliminary estimate. This does not exclude the possibility, that 
appears to us to be strongly motivated, that further theoretical 
and experimental efforts might reduce this value to a final limit
of, say, 15 or 10 \% size. In the final Section of this paper, we shall show
what might be the level of theoretical information obtainable under these 
 assumptions.

To conclude this Section, we want to discuss briefly the possibility that we 
mentioned  of using as realistic experimental input the quantity $A_t$, the final 
top longitudinal polarization asymmetry, defined in our Eq.~(\ref{def:at}), whose 
potential virtues were qualitatively investigated in Section 2. As we said in 
that Section, this quantity would exhibit the remarkable property of being 
essentially 
free of QCD effects (and, more important, of the related 
theoretical uncertainties). In our opinion, this would motivate a rigorous 
experimental analysis, like the one that we have shown for the unpolarized 
cross section. This analysis is not, at the moment, available. To provide a 
motivation for it, we shall show in the next Section the additional 
information that would be achievable if both measurements  of $\sigma_U$ and $A_t$ 
were performed. With this purpose, we shall simulate for $A_t$ three different 
overall errors of the same relative size (i.e. 20, 15, 10 \%) 
that we assumed for $\sigma_U$. Although we realize that these values might be 
unduly optimistic (or, possibly and hopefully, pessimistic), we
believe that the improvements of information that would be obtained should 
justify a dedicated experimental effort in that direction.

Section 3 is thus concluded. The following short Section 4 will be devoted to 
a review of the potential information that would be derivable from the results 
that we have listed, added to the formulae that we have derived in Section 2.

\section{Information on $\tan\beta$ from a realistic analysis of the large invariant 
mass dependence of top observables} 

In this final Section we shall show the results of the generalization of the  
$\chi^2$ analysis performed in Section 2, under the assumption of a purely 
statistical error, to the more realistic cases of an additional overall systematic 
(theoretical + experimental) relative error on 
the unpolarized cross section. The results of the fits are shown in Fig. 9. In the spirit of the 
last remarks of Section 3, we have also computed in Fig. 9 the results of 
three $\chi^2$ fits where, in addition to the unpolarized cross section, the 
longitudinal polarization asymmetry $A_t$ has also been included, assuming for it 
a relative error equal to that of the unpolarized cross section.

In more details, the left side of Fig.~(\ref{fig9}) shows the relative accuracy on 
$\tan\beta$ that can be derived
by exploiting both $\sigma_U$ and $A_t$ and adding to the statistical error a systematic
error constant in energy with three possible values, 10\%, 15\% and 20\%.
The right side of the Figure shows the same analysis, but without exploiting $A_t$.
From a comparison of the two plots, one sees that the contribution from $A_t$ is definitely
not negligible especially at larger values of the systematic error.
For instance, the determination of $\tan\beta$ using both $\sigma_U$ and $A_t$
with a systematic 20 \% error
gives an accuracy roughly equal to that obtained using only $\sigma_U$
with a systematic 15\% error. 

Apart from the specific role of $A_t$, we stress that the results shown in Fig.~(\ref{fig9})
are, in our opinion, encouraging. 
Indeed, even in the worst case, {\em i.e.} discarding $A_t$ and using 
only $\sigma_U$ at 20\%, 
we find an accuracy on $\tan\beta$ that is around 35\% at $\tan\beta \simeq 50$.
We emphasize again that this is due to the known special role of $\tan\beta$ appearing 
directly in the couplings and allowing a wide range of  enhancement in the radiative corrections.
Indeed, even with an error of 20\% on the experimental measurements, the radiative 
effects at $\tan\beta > 50$ can be large enough to be visible and able to 
constrain the possible values of $\tan\beta$ in the fit.

We also stress, to conclude this Section,  that the hypothesis of a systematic error of 20 \% 
seems to us indeed pessimistic since, for instance, an overall 5\% is the declared goal 
required for a determination of $m_t$ with a precision of 1 GeV (\cite{CERNYB}, page 428).

\section{Concluding remarks}

As a final comment to our work, we would like to make the following statement. 
The main purpose of our paper was to show that, at LHC, precision tests 
of  the Minimal Supersymmetric Standard Model would actually be possible,  
even if the experimental conditions will never be at the level of precision of 
LEP1, where the Standard Model was tested at the "less than one percent" level. 
The qualitative level of "optimal" accuracy would typically be here at the 
"ten percent" level, perhaps a little worse but possibly a little better. 
Under these conditions, we have shown that, even at an accuracy level of 
(only) twenty percent, the production of top-antitop pairs in a region of 
relatively large (but experimentally realistic) invariant mass could provide a 
non trivial confirmation (or, perhaps, a prediction) for the value of one of 
the most important parameters of the model, $\tan\beta$. The technical reasons of 
this possibility are that (a) this process contains contributions of Yukawa 
kind from vertices where the heavy top mass enters quadratically, enhanced by 
a large Sudakov logarithm and by quadratic factors of $\tan\beta$ and (b) that 
$\tan\beta$ is actually the \underline{only} SUSY parameter that enters the logarithmic 
expansion. The combination of these enhancements might in fact lead to an 
observable effect, in particular from a proper $\chi^2$ fit to the invariant mass 
distribution. Thus, from the accurate measurements of the production of the 
heaviest quark pair producible in the reaction, important tests of the model 
(or even predictions, in case $\tan\beta$ were still poorly known)  might be 
derived. We like to remark that
an analogy would exist with the analogous situation met at LEP1, where, from 
accurate measurements of the producion of $b\overline{b}$ pairs, fundamental tests and 
"predictions" for the value of the top mass were derived.

One should also (honestly) repeat, to conclude, that our (preliminary) 
analysis was performed assuming special conditions both for the model (a 
"moderately' light SUSY scenario) and for the logarithmic Sudakov expansion 
(next-to -next to leading order). In a less favourable case, though, our 
analysis could  be repeated with  the proper values of all the SUSY parameters 
and a complete one-loop description, which would not benefit from the simple 
Sudakov expansion. This more complete analysis in in fact already being 
carried on in~\cite{ToAppear}.

\newpage

\begin{figure}
\vspace{1.5cm}
\centering
\epsfig{file=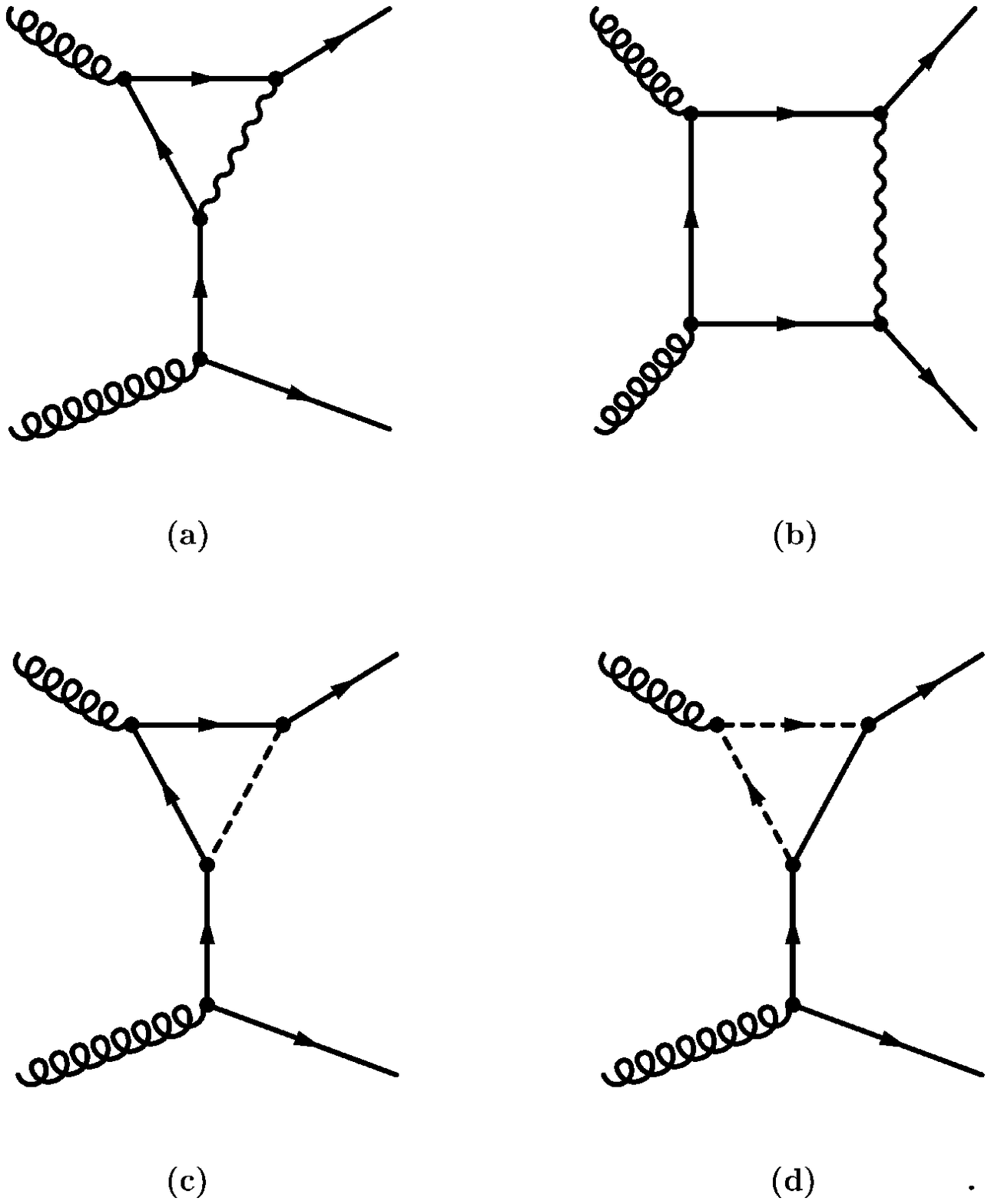,width=12cm}
\vspace{1.5cm}
\caption{Diagrams for electroweak Sudakov logarithmic corrections
to $gg\to t\bar t$. In (a),(b) the virtual gauge boson can be a
photon, a $Z$ or a $W^{\pm}$. In (c) the dashed line represents a
Higgs or a Goldstone particle ($H^{\pm,0}$, $h^0$, $A^0$,
$G^{\pm,0}$). In (d) the dashed lines represent a
sbottom or a stop and the solid line inside the triangle 
represents a chargino or a
neutralino.}
\label{diag}
\end{figure}

\begin{figure}
\centering
\epsfig{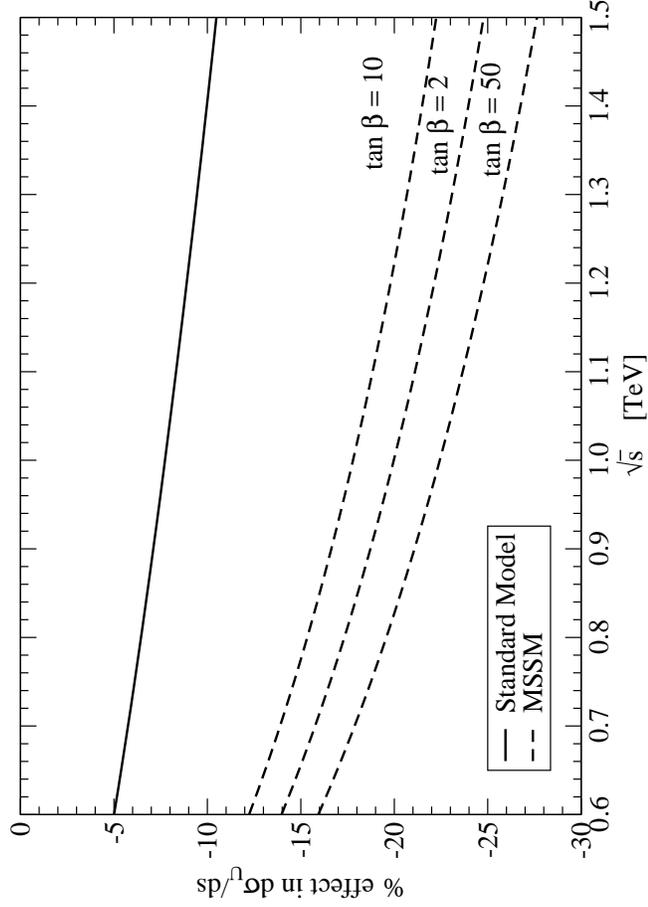}
\vspace{1.5cm}
\caption{Relative one loop corrections to the unpolarized cross section $d\sigma_U/ds$.
More precisely, we show the difference $d\sigma_U^{1\ loop}/ds-d\sigma_U^{Born}/ds$ divided by 
$d\sigma_U^{Born}/ds$.
The plot shows the effect in the Standard Model and in the MSSM with three reference
values of $\tan\beta = 2, 10, 50$.}
\label{fig2}
\end{figure}

\newpage

\begin{figure}
\centering
\epsfig{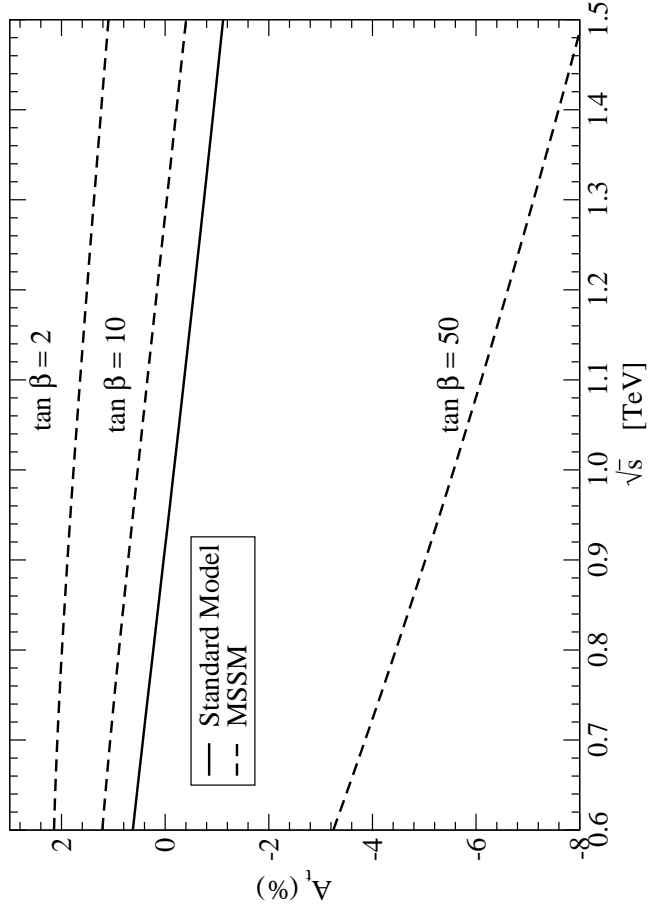}
\vspace{1.5cm}
\caption{One loop value of the top quark polarization asymmetry $A_t$.
The plot shows the effect in the Standard Model and in the MSSM with three reference
values of $\tan\beta = 2, 10, 50$.}
\label{fi}
\label{fig3}
\end{figure}

\newpage

\begin{figure}
\centering
\epsfig{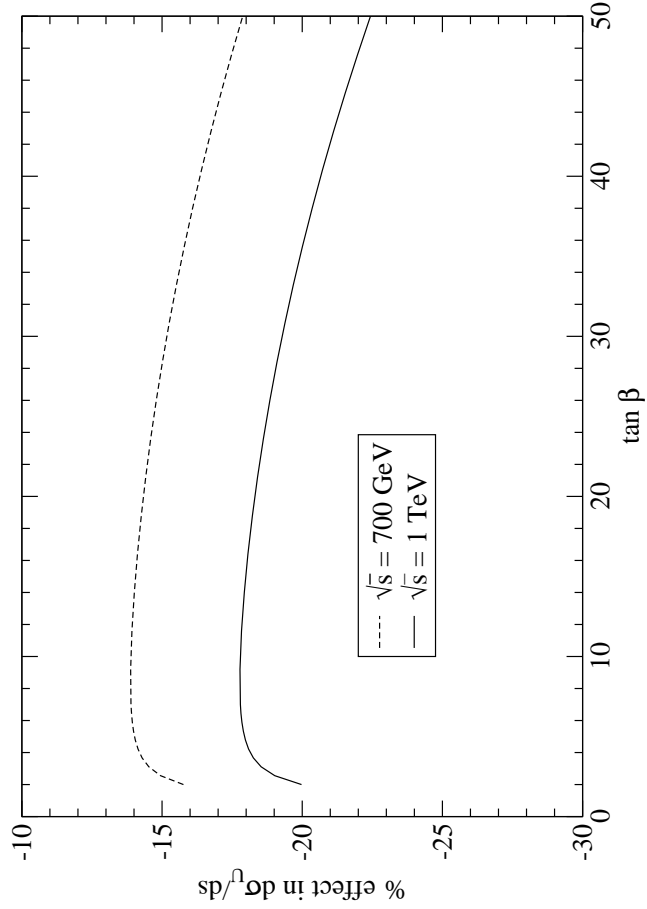}
\vspace{1.5cm}
\caption{Relative one loop corrections to the unpolarized cross section $d\sigma_U/ds$.
As in Fig.~\ref{fig2}, we show the ratio $(d\sigma_U^{1\ loop}/ds-d\sigma_U^{Born}/ds)/d\sigma_U^{Born}/ds$.
The plot shows the effect in the MSSM as a function of $\tan\beta$ at two reference
values of $\sqrt{s} = 700, 1000$ GeV.}
\label{fig4}
\end{figure}

\newpage

\begin{figure}
\centering
\epsfig{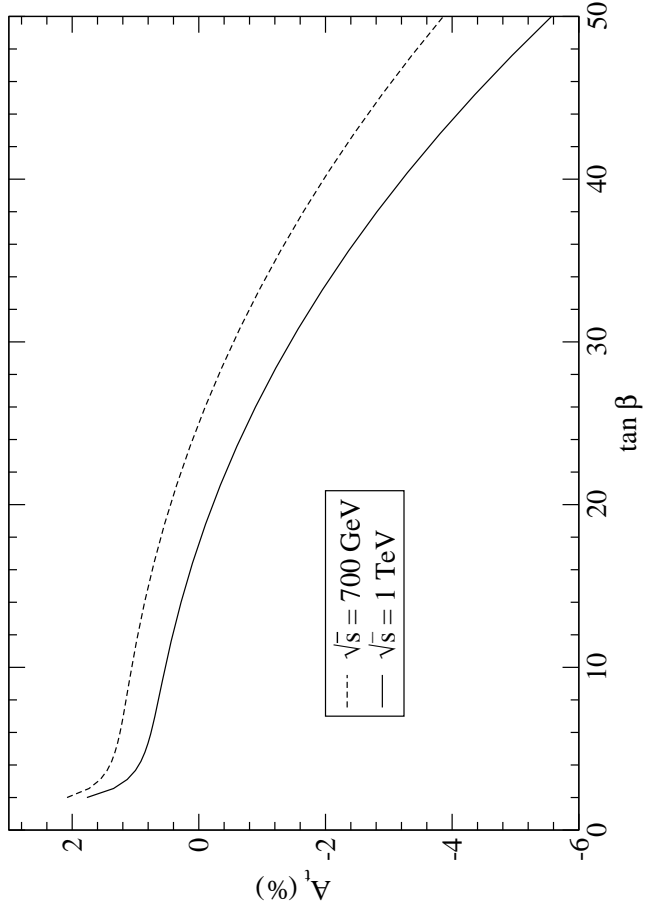}
\vspace{1.5cm}
\caption{One loop expression of the top quark polarization asymmetry $A_t$.
The plot shows the effect in the MSSM as a function of $\tan\beta$ at two reference
values of $\sqrt{s} = 700, 1000$ GeV.}
\label{fi}
\label{fig5}
\end{figure}

\newpage

\begin{figure}
\centering
\epsfig{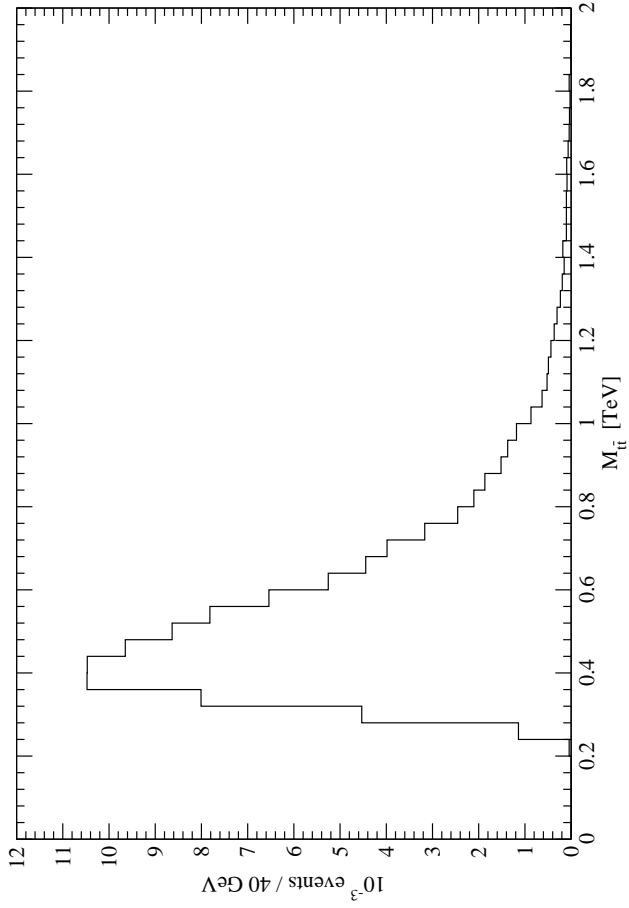}
\vspace{1.5cm}
\caption{Event distribution as function of $M_{t\bar t}$ in the Standard Model. The 
events are obtained from semileptonic top decay corresponding to 5 fb$^{-1}$ of
collected data (half a year  at luminosity $10^{33} {\rm cm}^{-2} {\rm s}^{-1}$).}
\label{fig6}
\end{figure}

\newpage

\begin{figure}
\centering
\epsfig{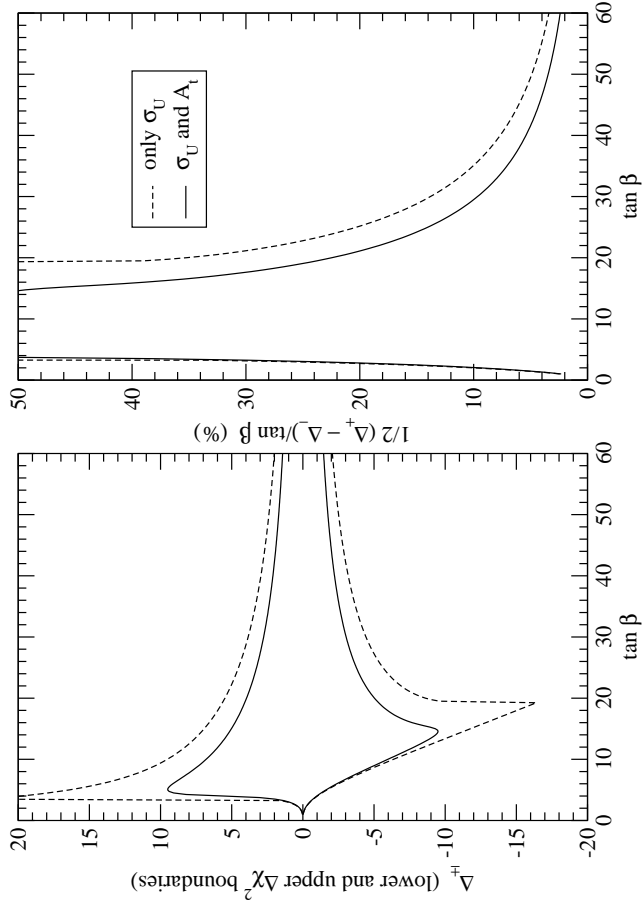}
\vspace{1.5cm}
\caption{Results from the $\chi^2$ analysis of the $\tan\beta$ dependence of the two 
considered distributions. For each hypothetical {\em true} value of 
$\tan\beta$, the figures show the $\Delta\chi^2=1$ boundaries $\tan\beta +\Delta_{\pm}$
and the corresponding relative error defined as $\frac{1}{2}(\Delta_+-\Delta-)/\tan\beta$.
The dashed and solid lines are obtained by exploiting the unpolarized cross section alone or
the combination of it with the polarization asymmetry.}
\label{fig7}
\end{figure}

\newpage

\begin{figure}
\centering
\epsfig{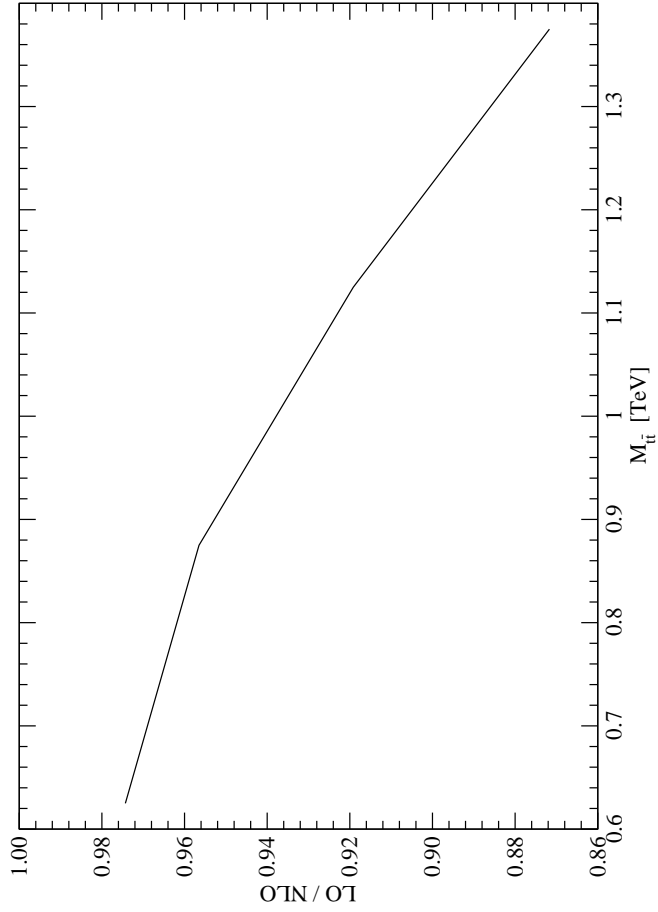}
\vspace{1.5cm}
\caption{Ratio of normalised NLO and LO predictions of the
distribution of $M_{tt}$, i.e. $[M_{tt}/\sum M_{tt}]/[\sqrt{s}/\sum \sqrt{s}]$, as determined from the
MC@NLO and Herwig programs. This ratio is only sensitive to the
difference in {\em shape} of the two predictions.
}
\label{fig8}
\end{figure}

\newpage

\begin{figure}
\centering
\epsfig{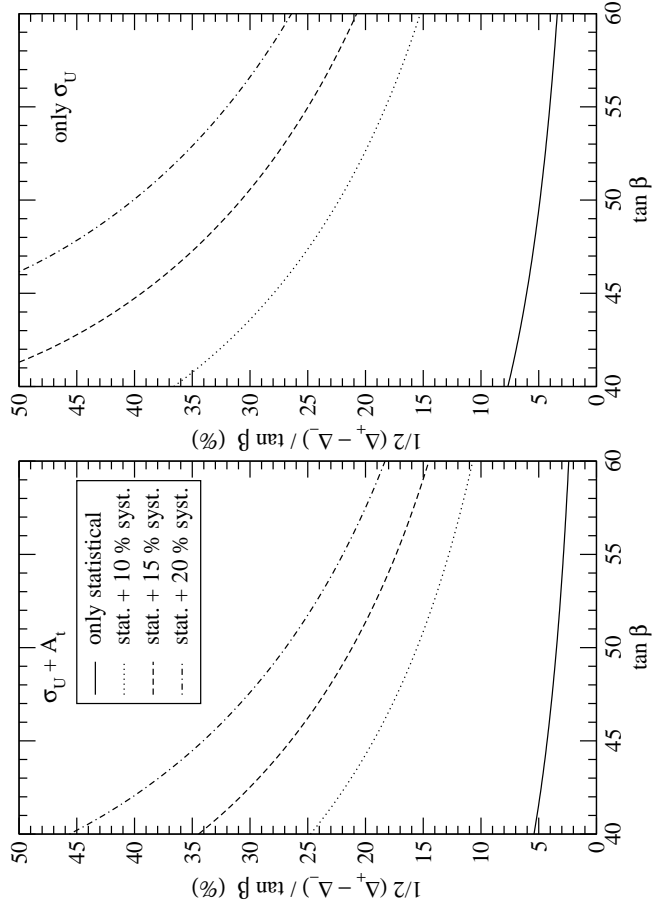}
\vspace{1.5cm}
\caption{This figure shows the same analysis described in Fig.~(\ref{fig7}).
However, we now take into account an energy independent fixed systematic
error of 10, 15 or 20 \%.
The curves on the left are obtained by including in the analysis both the unpolarized cross section and 
the polarization asymmetry $A_t$. The curves on the right do not use $A_t$.}
\label{fig9}
\end{figure}

\end{document}